\documentclass[12pt,a4paper]{article}

\usepackage{ifthen} 
\newboolean{pdflatex}
\setboolean{pdflatex}{true} 

\newboolean{articletitles}
\setboolean{articletitles}{true} 

\newboolean{uprightparticles}
\setboolean{uprightparticles}{false} 

\usepackage[compat=1.1.0]{tikz-feynman}

\def\paperauthors{LHCb collaboration} 
\def\paperasciititle{Observation of the doubly charmed baryon Xicc+
with the LHCb Run~3 detector} 
\def\papertitle{Observation of the doubly charmed baryon $\Xiccp$ with the LHCb Run~3 detector} 
\def\paperkeywords{{High Energy Physics}, {LHCb}} 
\def\papercopyright{\the\year\ CERN for the benefit of the LHCb collaboration} 
\def\paperlicence{CC BY 4.0 licence}
\def\paperlicenceurl{https://creativecommons.org/licenses/by/4.0/}

\newif\ifEnableSectionTOCLinks
\EnableSectionTOCLinksfalse 

\usepackage[top=1in, bottom=1.25in, left=1in, right=1in]{geometry}

\usepackage{microtype}
\usepackage{lineno}  
\usepackage{xspace} 
\usepackage{caption} 

\usepackage{graphicx}  
\usepackage{color}
\usepackage{colortbl}
\graphicspath{{./figs/}} 

\usepackage{amsmath} 
\usepackage{amssymb}
\usepackage{amsfonts}
\usepackage{upgreek} 

\newcommand*\patchAmsMathEnvironmentForLineno[1]{%
\expandafter\let\csname old#1\expandafter\endcsname\csname #1\endcsname
\expandafter\let\csname oldend#1\expandafter\endcsname\csname
end#1\endcsname
 \renewenvironment{#1}%
   {\linenomath\csname old#1\endcsname}%
   {\csname oldend#1\endcsname\endlinenomath}%
}
\newcommand*\patchBothAmsMathEnvironmentsForLineno[1]{%
  \patchAmsMathEnvironmentForLineno{#1}%
  \patchAmsMathEnvironmentForLineno{#1*}%
}
\AtBeginDocument{%
\patchBothAmsMathEnvironmentsForLineno{equation}%
\patchBothAmsMathEnvironmentsForLineno{align}%
\patchBothAmsMathEnvironmentsForLineno{flalign}%
\patchBothAmsMathEnvironmentsForLineno{alignat}%
\patchBothAmsMathEnvironmentsForLineno{gather}%
\patchBothAmsMathEnvironmentsForLineno{multline}%
\patchBothAmsMathEnvironmentsForLineno{eqnarray}%
}

\usepackage[pdftex,
            pdfauthor={\paperauthors},
            pdftitle={\paperasciititle},
            pdfkeywords={\paperkeywords}]{hyperref}
\usepackage{hyperxmp}
\hypersetup{
    pdfcopyright={Copyright (C) \papercopyright},
    pdflicenseurl={\paperlicenceurl}
}

\usepackage[colorinlistoftodos,textsize=scriptsize]{todonotes}

\usepackage[bottom,flushmargin,hang,multiple]{footmisc}

\usepackage[all]{hypcap} 

\usepackage{xspace} 
\usepackage{upgreek}

\def\lhcb   {\mbox{LHCb}\xspace}

\def\babar  {\mbox{BaBar}\xspace}
\def\belle  {\mbox{Belle}\xspace}

\def\MagUp {\mbox{\em Mag\kern -0.05em Up}\xspace}

\ifthenelse{\boolean{uprightparticles}}%
{

 \def\Pmu         {\ensuremath{\upmu}\xspace}

 \def\Ppi         {\ensuremath{\uppi}\xspace}

 \def\Ppsi        {\ensuremath{\uppsi}\xspace}

 \def\PDelta      {\ensuremath{\Delta}\xspace}                 
 \def\PXi         {\ensuremath{\Xi}\xspace}                 
 \def\PLambda     {\ensuremath{\Lambda}\xspace}                 
 \def\PSigma      {\ensuremath{\Sigma}\xspace}                 
 \def\POmega      {\ensuremath{\Omega}\xspace}                 
 \def\PUpsilon    {\ensuremath{\Upsilon}\xspace}
 \let\oldPi\Pi
 \def\PPi         {\ensuremath{\oldPi}\xspace}

 \def\PB      {\ensuremath{\mathrm{B}}\xspace}                 
 \def\PD      {\ensuremath{\mathrm{D}}\xspace}                 
 \def\PJ      {\ensuremath{\mathrm{J}}\xspace}                 
 \def\PK      {\ensuremath{\mathrm{K}}\xspace}                 
 \def\PW      {\ensuremath{\mathrm{W}}\xspace}                 
 \def\Pb      {\ensuremath{\mathrm{b}}\xspace}                 
 \def\Pc      {\ensuremath{\mathrm{c}}\xspace}                 
 \def\Pd      {\ensuremath{\mathrm{d}}\xspace}                 

 \def\Pp      {\ensuremath{\mathrm{p}}\xspace}                 

 \def\Ps      {\ensuremath{\mathrm{s}}\xspace}                 
                  
 \def\Pu      {\ensuremath{\mathrm{u}}\xspace}                 
 \def\thebaroffset{0.0em}
}
{

 \def\Pmu         {\ensuremath{\mu}\xspace}

 \def\Ppi         {\ensuremath{\pi}\xspace}

 \def\Ppsi        {\ensuremath{\psi}\xspace}                 
                  
 \mathchardef\PDelta="7101
 \mathchardef\PXi="7104
 \mathchardef\PLambda="7103
 \mathchardef\PSigma="7106
 \mathchardef\POmega="710A
 \mathchardef\PUpsilon="7107
 \mathchardef\PPi="7105
 \def\PB      {\ensuremath{B}\xspace}                 
 \def\PD      {\ensuremath{D}\xspace}                 
 \def\PJ      {\ensuremath{J}\xspace}                 
 \def\PK      {\ensuremath{K}\xspace}                 
 \def\PW      {\ensuremath{W}\xspace}                 
 \def\Pb      {\ensuremath{b}\xspace}                 
 \def\Pc      {\ensuremath{c}\xspace}                 
 \def\Pd      {\ensuremath{d}\xspace}                 

 \def\Pp      {\ensuremath{p}\xspace}                 

 \def\Ps      {\ensuremath{s}\xspace}                 
                  
 \def\Pu      {\ensuremath{u}\xspace}                 
 \def\thebaroffset{0.18em}
}
\newcommand{\offsetoverline}[2][\thebaroffset]{\kern #1\overline{\kern -#1 #2}}%

\makeatletter
\ifcase \@ptsize \relax
  \newcommand{\miniscule}{\@setfontsize\miniscule{4}{5}}
\or
  \newcommand{\miniscule}{\@setfontsize\miniscule{5}{6}}
\or
  \newcommand{\miniscule}{\@setfontsize\miniscule{5}{6}}
\fi
\makeatother

\DeclareRobustCommand{\optbar}[1]{\shortstack{{\miniscule (\rule[.5ex]{1.25em}{.18mm})}
  \\ [-.7ex] $#1$}}

\def\mumu       {{\ensuremath{\Pmu^+\Pmu^-}}\xspace}

\def\W      {{\ensuremath{\PW}}\xspace}

\def\uquark    {{\ensuremath{\Pu}}\xspace}

\def\dquark    {{\ensuremath{\Pd}}\xspace}

\def\squark    {{\ensuremath{\Ps}}\xspace}

\def\cquark    {{\ensuremath{\Pc}}\xspace}

\def\bquark    {{\ensuremath{\Pb}}\xspace}

\def\pion   {{\ensuremath{\Ppi}}\xspace}

\def\pip    {{\ensuremath{\pion^+}}\xspace}
\def\pim    {{\ensuremath{\pion^-}}\xspace}

\def\kaon    {{\ensuremath{\PK}}\xspace}

\def\KorKbar {\kern \thebaroffset\optbar{\kern -\thebaroffset \PK}{}\xspace}

\def\Kp      {{\ensuremath{\kaon^+}}\xspace}
\def\Km      {{\ensuremath{\kaon^-}}\xspace}

\def\D       {{\ensuremath{\PD}}\xspace}

\def\DorDbar {\kern \thebaroffset\optbar{\kern -\thebaroffset \PD}\xspace}

\def\Dp      {{\ensuremath{\D^+}}\xspace}
\def\Dm      {{\ensuremath{\D^-}}\xspace}

\def\DpDm    {\ensuremath{\Dp {\kern -0.16em \Dm}}\xspace}

\def\B       {{\ensuremath{\PB}}\xspace}

\def\BorBbar {\kern \thebaroffset\optbar{\kern -\thebaroffset \PB}\xspace}

\def\Bd      {{\ensuremath{\B^0}}\xspace}

\def\BdorBdbar {\kern \thebaroffset\optbar{\kern -\thebaroffset \Bd}\xspace}
\def\Bu      {{\ensuremath{\B^+}}\xspace}

\def\Bs      {{\ensuremath{\B^0_\squark}}\xspace}

\def\BsorBsbar {\kern \thebaroffset\optbar{\kern -\thebaroffset \Bs}\xspace}

\def\jpsi     {{\ensuremath{{\PJ\mskip -3mu/\mskip -2mu\Ppsi}}}\xspace}

\def\Y#1S{\ensuremath{\PUpsilon{(#1S)}}\xspace}

\def\proton      {{\ensuremath{\Pp}}\xspace}

\def\Lz          {{\ensuremath{\PLambda}}\xspace}

\def\LorLbar     {\kern \thebaroffset\optbar{\kern -\thebaroffset \PLambda}\xspace}

\def\Xires       {{\ensuremath{\PXi}}\xspace}

\def\Lc          {{\ensuremath{\Lz^+_\cquark}}\xspace}

\def\Xicp        {{\ensuremath{\Xires^+_\cquark}}\xspace}

\def\Xiccp       {{\ensuremath{\Xires^+_{\cquark\cquark}}}\xspace}
\def\Xiccpp      {{\ensuremath{\Xires^{++}_{\cquark\cquark}}}\xspace}

\newcommand{\decay}[2]{\ensuremath{\mathinner{#1\!\to #2}}\xspace}

\def\to                 {\ensuremath{\rightarrow}\xspace}

\def\AT#1     {\ensuremath{A_{\mathrm{T}}^{#1}}\xspace}           

\def\C#1      {\ensuremath{\mathcal{C}_{#1}}\xspace}                       
\def\Cp#1     {\ensuremath{\mathcal{C}_{#1}^{'}}\xspace}                    
\def\Ceff#1   {\ensuremath{\mathcal{C}_{#1}^{\mathrm{(eff)}}}\xspace}        
\def\Cpeff#1  {\ensuremath{\mathcal{C}_{#1}^{'\mathrm{(eff)}}}\xspace}       
\def\Ope#1    {\ensuremath{\mathcal{O}_{#1}}\xspace}                       
\def\Opep#1   {\ensuremath{\mathcal{O}_{#1}^{'}}\xspace}                    

\newcommand{\aunit}[1]{\ensuremath{\text{\,#1}}}       

\newcommand{\tev}{\aunit{Te\kern -0.1em V}\xspace}
\newcommand{\gev}{\aunit{Ge\kern -0.1em V}\xspace}
\newcommand{\mev}{\aunit{Me\kern -0.1em V}\xspace}
\newcommand{\kev}{\aunit{ke\kern -0.1em V}\xspace}
\newcommand{\ev}{\aunit{e\kern -0.1em V}\xspace}
 
\newcommand{\mevc}{\ensuremath{\aunit{Me\kern -0.1em V\!/}c}\xspace}
\newcommand{\gevc}{\ensuremath{\aunit{Ge\kern -0.1em V\!/}c}\xspace}
\newcommand{\mevcc}{\ensuremath{\aunit{Me\kern -0.1em V\!/}c^2}\xspace}
\newcommand{\gevcc}{\ensuremath{\aunit{Ge\kern -0.1em V\!/}c^2}\xspace}

\def\fb   {\ensuremath{\aunit{fb}}\xspace}
\def\invfb   {\ensuremath{\fb^{-1}}\xspace}

\def\fs   {\aunit{fs}}

\newcommand{\chisq}{\ensuremath{\chi^2}\xspace}

\def\gsim{{~\raise.15em\hbox{$>$}\kern-.85em
          \lower.35em\hbox{$\sim$}~}\xspace}
\def\lsim{{~\raise.15em\hbox{$<$}\kern-.85em
          \lower.35em\hbox{$\sim$}~}\xspace}

\def\sPlot{\mbox{\em sPlot}\xspace}

\def\pt         {\ensuremath{p_{\mathrm{T}}}\xspace}

\def\evtgen     {\mbox{\textsc{EvtGen}}\xspace}

\def\geant      {\mbox{\textsc{Geant4}}\xspace}

\def\photos     {\mbox{\textsc{Photos}}\xspace}

\def\pythia     {\mbox{\textsc{Pythia}}\xspace}

\def\tell1  {TELL1\xspace}
\def\ukl1   {UKL1\xspace}

\newcommand{\eg}{\mbox{\itshape e.g.}\xspace}

\newcommand{\lhcborcid}[1]{\href{https://orcid.org/#1}{\hspace*{0.1em}\raisebox{-0.45ex}{\includegraphics[width=1em]{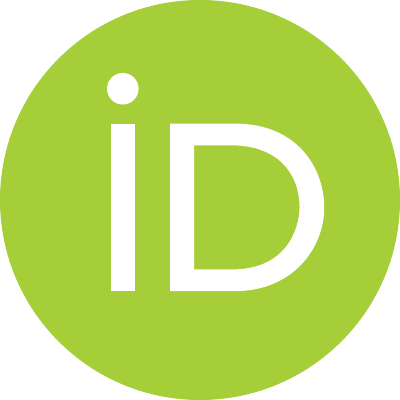}}}}

\hypersetup{
  colorlinks   = true, 
  urlcolor     = blue, 
  linkcolor    = blue, 
  citecolor    = red   
}

\ifEnableSectionTOCLinks
    \usepackage[explicit]{titlesec} 
    
    \let\oldcontentsline\contentsline
    \renewcommand

    \titleformat{\section}{\normalfont\Large\bf}{\hyperlink{tocsection.\thesection}{{\thesection} \parbox[t]{\dimexpr\textwidth-1pc}{#1}}}{1pc}{}

    \titleformat{\subsection}{\normalfont\bf}{\hyperlink{tocsubsection.\thesubsection}{{\thesubsection} \parbox[t]{\dimexpr\textwidth-1pc}{#1}}}{1pc}{}

    \titleformat{name=\section,numberless}[display]{}{}{0pt}{\normalfont\Huge\bfseries #1}
\fi

\usepackage{cite} 
\usepackage{LHCb/mciteplus}

\usepackage{longtable} 
\def\totLumi   {\ensuremath{6.9\mbox{\,fb}^{-1}}\xspace}

\def\Xiccmass{{\ensuremath{3619.97 \pm 0.83 \pm 0.26 \,^{+1.90}_{-1.30} \mevcc}}\xspace}

\def\DeltaMass {{\ensuremath{-1.77 \pm 0.84 \pm 0.15 \,^{+1.90}_{-1.30}\mevcc}}\xspace}

\begin{document}

\renewcommand{\thefootnote}{\fnsymbol{footnote}}
\setcounter{footnote}{1}


\begin{titlepage}
\pagenumbering{roman}

\vspace*{-1.5cm}
\centerline{\large EUROPEAN ORGANIZATION FOR NUCLEAR RESEARCH (CERN)}
\vspace*{1.5cm}
\noindent
\begin{tabular*}{\linewidth}{lc@{\extracolsep{\fill}}r@{\extracolsep{0pt}}}
\ifthenelse{\boolean{pdflatex}}
{\vspace*{-1.5cm}\mbox{\!\!\!\includegraphics[width=.14\textwidth]{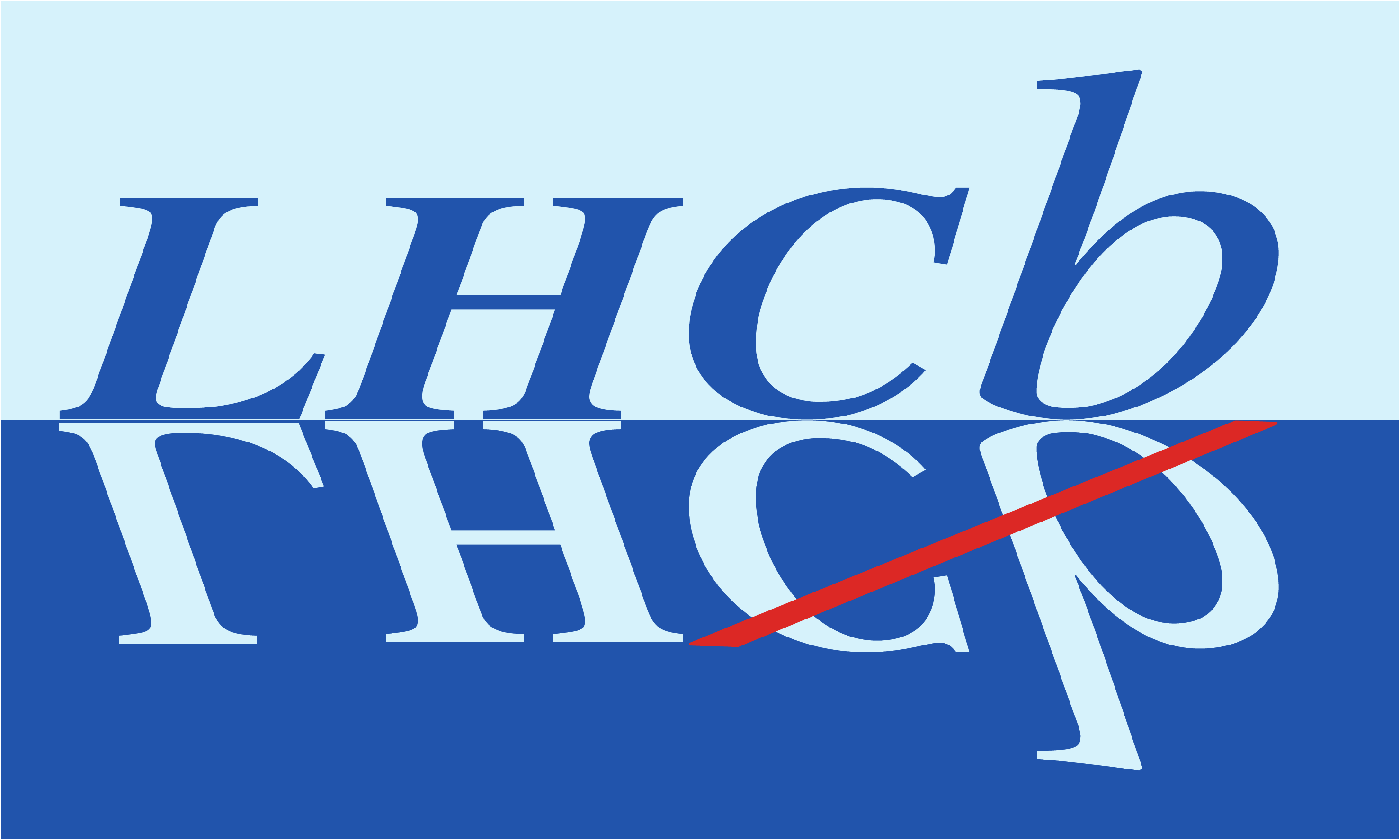}} & &}%
{\vspace*{-1.2cm}\mbox{\!\!\!\includegraphics[width=.12\textwidth]{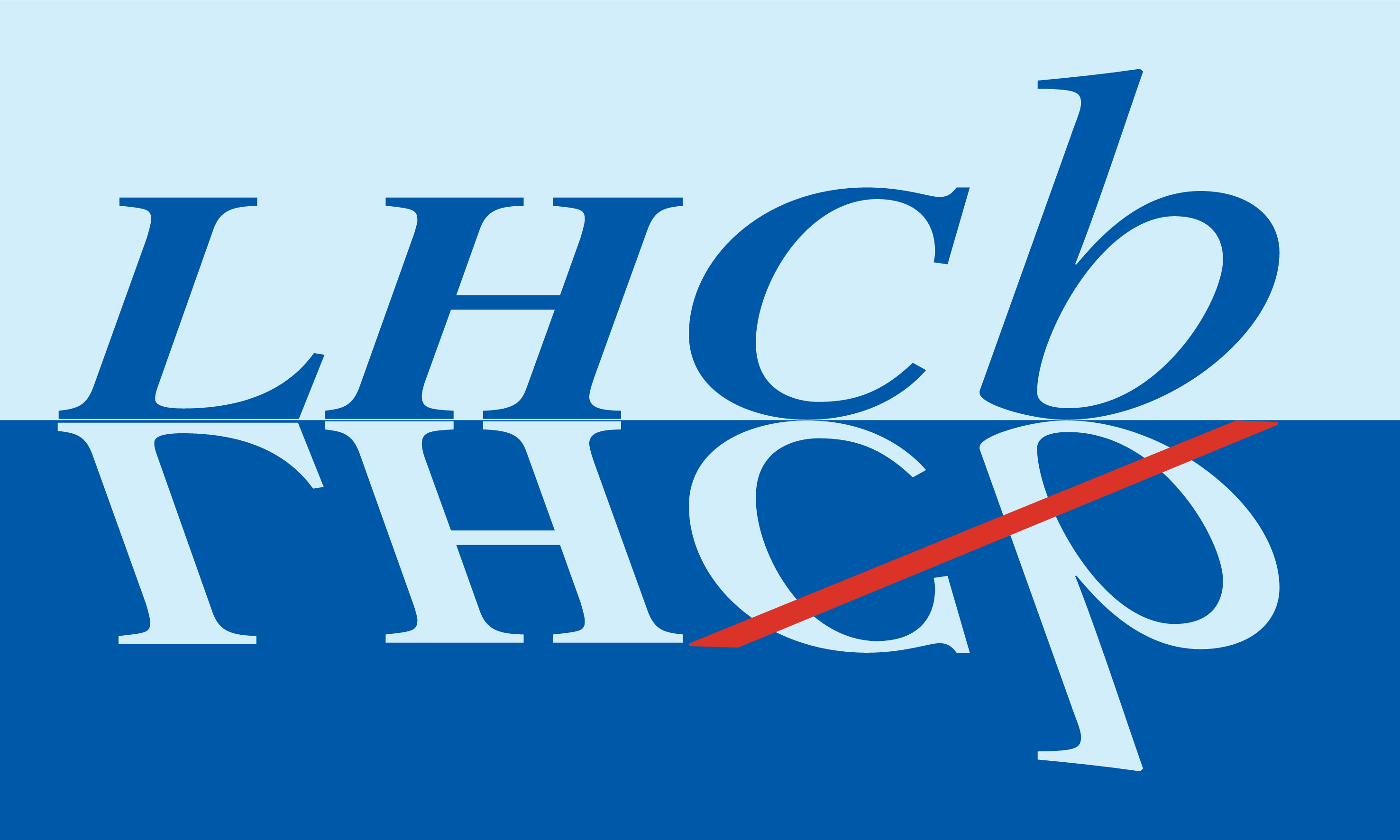}} & &}%
\\
 & & CERN-EP-2026-085 \\  
 & & LHCb-PAPER-2026-009 \\  
 & & July 8, 2026 \\ 
\end{tabular*}

\vspace*{4.0cm}

{\normalfont\bfseries\boldmath\huge
\begin{center}
  \papertitle 
\end{center}
}

\vspace*{2.0cm}

\begin{center}
\paperauthors\footnote{Authors are listed at the end of this Letter.}
\end{center}

\vspace{\fill}

\begin{abstract}
  \noindent
  The first observation of the doubly charmed baryon 
  \Xiccp is reported through
  its decay to the $\Lc\Km\pip$ final state, with a statistical significance exceeding 
  seven standard deviations.  
  The observation is made 
  using proton-proton collision data collected in 2024 with the \lhcb Run~3 detector 
  at a center-of-mass energy of 13.6\tev, corresponding to a total integrated luminosity of \totLumi.
  The $\Xiccp$ mass is measured to be \Xiccmass,
  where the first uncertainty is statistical, the second is systematic, and the third 
  is due to the unknown $\Xiccp$ lifetime, 
  which is assumed to lie in the range 15--160$\fs$ with a baseline value of 45\fs. 
  The difference between the masses of the $\Xiccp$ and $\Xiccpp$ baryons is determined to be $\mbox{\DeltaMass}$. 
  This is the first observation of a new particle made with the 
  \lhcb Run~3 detector.

\end{abstract}

\vspace*{2.0cm}

\begin{center}
  Published in \href{https://doi.org/10.1103/dmv6-7gdv}{Phys.~Rev.~Lett. 137 (2026) 021902} 
\end{center}

\vspace{\fill}

{\footnotesize 
\centerline{\copyright~\papercopyright. \href{\paperlicenceurl}{\paperlicence}.}}
\vspace*{2mm}

\end{titlepage}


\newpage
\setcounter{page}{2}
\mbox{~}


\renewcommand{\thefootnote}{\arabic{footnote}}
\setcounter{footnote}{0}


\cleardoublepage


\pagestyle{plain} 
\setcounter{page}{1}
\pagenumbering{arabic}


In the constituent quark model~\cite{
  GellMann:1964nj,
  Zweig:352337,
  *Zweig:570209}, 
the $\Xiccp$ baryon is the isospin partner 
to the $\Xiccpp$ baryon (with quark content $\Xiccp=\cquark\cquark\dquark$ and $\Xiccpp=\cquark\cquark\uquark$),
just like the neutron ($\uquark\dquark\dquark$) to the proton ($\uquark\uquark\dquark$), 
and therefore has approximately the same production cross-section 
as the well-established $\Xiccpp$ baryon~\cite{Berezhnoy:1998aa,
Kiselev:2001fw,Ma:2003zk,Chang:2006xp,Chang:2006eu,
Zhang:2011hi,
Chang:2005bf,
Sun:2020mvl,
Yang:2024ysg,
Zhao:2025kvq}. 
Unlike the neutron, which is more massive than the proton by 
about $1.3\mevcc$~\cite{Tiesinga:2021myr}, the $\Xiccp$ baryon is expected to have a smaller mass 
than the $\Xiccpp$ baryon by a few $\!\mevcc$, 
due to the competition between electromagnetic and isospin-breaking effects~\cite{Hwang:2008dj,
Brodsky:2011zs,
BMW:2014pzb,
Wei:2015gsa,
Karliner:2017gml}.
The masses of both $\Xiccp$ and $\Xiccpp$ baryons 
have been predicted by 
various theoretical models~\cite{Kerbikov:1987vx,
  Fleck:1989mb,
  Korner:1994nh,
  Chernyshev:1995gj,
  Ebert:1996ec, Gershtein:1998sx,*Gershtein:1998un,*Gershtein:2000nx,
  Itoh:2000um,
  Anikeev:2001rk,
  Kiselev:2001fw,
  Ebert:2002ig,
  Mathur:2002ce,
  He:2004px,
  Roberts:2007ni,
  Valcarce:2008dr,
  Hwang:2008dj,
  Zhang:2008rt,
  Wang:2010hs,
  Brodsky:2011zs,
  Aliev:2012ru,
  Namekawa:2013vu,
  BMW:2014pzb,
  Karliner:2014gca,
  Brown:2014ena,
  Sun:2014aya,
  Padmanath:2015jea,
  Wei:2015gsa,
  Bali:2015lka,
  Sun:2016wzh,
  Alexandrou:2017xwd,
  Karliner:2017gml,
  Liu:2017xzo,*Liu:2017frj,
  Ghalenovi:2022dok,
  Yu:2022lel,
  Pan:2025lmd,
  Jakhad:2025zos},
  and the $\Xiccpp$ mass has been measured to be 
  $3621.55 \pm 0.38\mevcc$~\cite{LHCb-PAPER-2019-037}.
The $\Xiccp$ lifetime is predicted to be three to six times
shorter than the $\Xiccpp$ lifetime,  
due to the \mbox{$\W$-exchange} contribution in the \Xiccp decay
and the constructive (destructive) Pauli interference of the $\Xiccp\; (\Xiccpp)$ decay amplitudes~\cite{Fleck:1989mb,
  guberina1999inclusive,
  Kiselev:1998sy,
  Likhoded:1999yv,
  Onishchenko:2000yp,
  Anikeev:2001rk,
  Kiselev:2001fw,
  hsi2008lifetime,
  Karliner:2014gca,
  Berezhnoy2016,
  Cheng:2018mwu,
  Dulibic:2023jeu}.
    After considering the measured $\Xiccpp$ lifetime to be
  $256 \pm 27\fs$~\cite{LHCb-PAPER-2018-019}, 
   the $\Xiccp$ lifetime is assumed to lie
   in the range \mbox{36--138$\fs$}.
The first theoretical prediction accounting for higher-order
contributions gives a $\Xiccp$ lifetime of $45\fs$~\cite{Cheng:2018mwu},
and this value is used as the baseline for the simulation in this analysis.

The $\Xiccp$ baryon has been searched for by
several experiments in the past decades. 
The SELEX collaboration claimed an observation of the \Xiccp baryon decaying into the \Lc\Km\pip and $\Dp \proton \Km$ final states~\cite{
  Mattson:2002vu,
  Ocherashvili:2004hi},
using a 600\gevc charged hyperon beam impinging on a fixed target.
The mass of the \Xiccp baryon, averaged over the two decay modes, was found to be \mbox{$3518.7 \pm 1.7 \mevcc$}, and the lifetime was measured to be less than $33\fs$ at the 90\% confidence level.
It was estimated that about 20\% of \Lc baryons in the SELEX experiment were produced from \Xiccp decays~\cite{Mattson:2002vu}.
Searches in different production environments
by the FOCUS~\cite{Ratti:2003ez},
\babar~\cite{Aubert:2006qw},
\lhcb~\cite{LHCb-PAPER-2013-049,LHCb-PAPER-2019-029,LHCb-PAPER-2021-019}
and \belle~\cite{Kato:2013ynr} collaborations
did not confirm the SELEX results. 
Meanwhile,   
the \Xiccpp baryon was observed by the LHCb experiment
in the $\Lc\Km\pip\pip$ final state~\cite{LHCb-PAPER-2017-018},
and confirmed in the $\Xicp \pip$ final
state~\cite{LHCb-PAPER-2018-026}.
The $\Xiccpp$ mass was measured in these decay modes to be 
about $100\mevcc$ higher than the mass of the $\Xiccp$ baryon claimed by SELEX, in contradiction to the theoretical expectation noted above. Recently, the LHCb collaboration 
has observed two additional decay modes of the $\Xiccpp$ baryon, 
namely \decay{\Xiccpp}{\PXi_c^{\prime+} \pi^+}~\cite{LHCb-PAPER-2021-052}
and 
\decay{\Xiccpp}{\PXi_c^0 \pi^+\pi^+}~\cite{LHCb-PAPER-2024-053}.

This Letter presents a 
search for the $\Xiccp$ baryon 
via the decay mode \decay{\Xiccp}{\Lc\Km\pip}, 
where the $\Lc$ baryon is reconstructed in the final state $\proton\Km\pip$. 
An example Feynman diagram of the decay is shown in Fig.~\ref{fig:feynman}. 
The inclusion of charge-conjugate decay processes is implied throughout this Letter. 
The dataset comprises proton-proton ($pp$) collision data 
recorded with the \lhcb Run~3 detector in 2024 at a center-of-mass energy
$\sqrt{s}=13.6\tev$,
corresponding to an integrated luminosity of \totLumi.
To avoid experimenter bias, the results of the analysis were
not examined until the full procedure had been finalized. 
The \decay{\Xiccpp}{\Lc\Km\pip\pip} decay is used as a control mode to develop the event-selection strategy and to help calibrate a precision mass measurement.

\begin{figure}[htbp]
    \centering
    \begin{tikzpicture}
        \begin{feynman}
        \vertex (c1) {\(c\)};
        \vertex[right=2cm of c1] (c2);
        \vertex[right=2cm of c2] (c3) {\(s\)};
        \vertex[above=5em of c1] (b1) {\(c\)};
        \vertex[above=5em of c3] (b2) {\(c\)};
        \vertex[above=1.5em of b1] (a1) {\(d\)};
        \vertex[above=1.5em of b2] (a2) {\(d\)};
        \vertex[below=2em of c3] (d1) {\(u\)};
        \vertex[below=1.5em of d1] (d3) {\(\overline d\)};
        \vertex[below=1.5em of b2] (e1) {\(u\)};
        \vertex[below=2em of e1] (e3) {\(\overline u\)};
        \vertex at ($(d1)!0.5!(d3) - (1cm, 0)$) (d2);
        \vertex at ($(e1)!0.5!(e3) - (1cm, 0)$) (e2);
        \diagram* {
        (a1) -- [fermion] (a2),
        (b1) -- [fermion] (b2),
        (c1) -- [fermion] (c2) -- [fermion] (c3),
        (d3) -- [fermion, out=180, in=-45] (d2) -- [fermion, out=45, in=180] (d1),
        (e3) -- [fermion, out=180, in=-45] (e2) -- [fermion, out=45, in=180] (e1),
        (c2) -- [boson, bend right, edge label=\(W^{+}\)] (d2),
        };
        \draw [decoration={brace}, decorate] (c1.south west) -- (a1.north west)
        node [pos=0.5, left] {\(\Xiccp\)};
        \draw [decoration={brace}, decorate] (a2.north east) -- (e1.south east)
        node [pos=0.5, right] {\(\Lc\)};
        \draw [decoration={brace}, decorate] (d1.north east) -- (d3.south east)
        node [pos=0.5, right] {\(\pip\)};
        \draw [decoration={brace}, decorate] (e3.north east) -- (c3.south east)
        node [pos=0.5, right] {\(\Km\)};
        \end{feynman}
    \end{tikzpicture}
    \caption{Example Feynman diagram contributing to the decay \decay{\Xiccp}{\Lc\Km\pip}.}
    \label{fig:feynman}
\end{figure}

The \lhcb Run~3  detector~\cite{LHCb-DP-2022-002} is a
single-arm forward spectrometer covering the pseudo\-rapidity range $2<\eta <5$, designed for 
the study of particles containing \bquark or \cquark quarks. 
Installed prior to the start of Run~3 data taking in 2022, it represents a major upgrade of the detector. The new configuration
has been designed to match the performance of the Run~1--2 detector~\cite{LHCb-DP-2022-002,LHCb-DP-2008-001}, while allowing it to operate 
at approximately five times the instantaneous
luminosity. 
The detector elements that are particularly
relevant to this analysis include: a silicon-pixel vertex detector surrounding the $pp$ interaction
region~\cite{LHCb-TDR-013} that allows 
hadrons containing beauty or charm 
quarks to be identified from their characteristically long
flight distances; a tracking system~\cite{LHCb-TDR-015} that provides a measurement of the momentum, $p$, of charged
particles; and two ring-imaging Cherenkov (RICH) detectors~\cite{LHCb-TDR-014} that are able to discriminate between
different species of charged hadrons. 
The readout of all detectors at the LHC collision frequency and the subsequent processing of the data
in a full-software trigger~\cite{LHCb-TDR-016, LHCb-TDR-017, LHCb-TDR-018, LHCb-TDR-021}
is a central feature of the upgraded detector, enabling the reconstruction and selection of events in real time. Compared with Run~2, 
the trigger efficiency of hadronic decay modes 
in the 2024 data taking has been improved by 
a factor of 2 to 4, depending on the number of final-state particles, 
thanks to the removal of the hardware trigger. 
The online reconstruction incorporates near-real-time alignment and calibration of the
detector~\cite{LHCb-PROC-2015-011}, which in turn allows the reconstruction of the $\PXi_{cc}^{+(+)}$
decays to be performed entirely by the trigger software.
The momentum scale is calibrated using samples of $\decay{\jpsi}{\mumu}$ 
and $\decay{\Bu}{\jpsi\Kp}$~decays collected concurrently
with the~data sample used for this analysis~\cite{LHCb-PAPER-2012-048,LHCb-PAPER-2013-011,LHCb-DP-2023-003},
and the calibration is applied offline along with the latest detector alignments. 
The~relative accuracy of this
procedure is estimated to be $2 \times 10^{-4}$ using samples of 
other fully reconstructed heavy-flavor hadrons including $\PUpsilon$~and
$\psi(2S)$~mesons.

Simulated samples of signal and control modes are required to model the effects of the detector acceptance and the imposed selection requirements.
  In the simulation, $pp$ collisions are generated using
  \pythia8~\cite{Sjostrand:2007gs,*Sjostrand:2006za} 
  with a specific \lhcb configuration~\cite{LHCb-PROC-2010-056}.
A dedicated generator, $\textsc{GenXicc}$~2.0~\cite{Chang:2009va}, is used to simulate the doubly charmed baryon production,
and is interfaced to \pythia8 to provide the underlying event, parton shower, and hadronization generation. 
Decays of unstable particles are described by \evtgen~\cite{Lange:2001uf}, in which final-state radiation is generated using \photos~\cite{davidson2015photos}. The interaction of the generated particles with the detector, and its response, are simulated using the \geant toolkit~\cite{Allison:2006ve, *Agostinelli:2002hh} 
 as described in Ref.~\cite{LHCb-PROC-2011-006}. Simulated 
 \decay{\Xiccp}{\Lc\Km\pip} (\decay{\Xiccpp}{\Lc\Km\pip\pip})
 samples are generated with a mass of $3621.4\mevcc$ and a lifetime of $45\fs$~\cite{
  Cheng:2018mwu,
  Dulibic:2023jeu} (256\fs~\cite{LHCb-PAPER-2018-019}), and samples with different lifetime hypotheses are obtained using a weighting technique. Both $\Xiccp$ and $\Xiccpp$ baryon decays are generated uniformly within the allowed phase space.

Candidate \decay{\Lc}{p\Km \pip} decays 
are reconstructed from three charged particles
identified as a proton, kaon and pion
using information from the RICH detectors. The particles are required to  
form a good-quality vertex and be inconsistent with originating from any 
primary $pp$ collision vertex (PV).
The $\Lc$ candidate is required to 
have a mass in the range from 2270 to $2306\mevcc$,
which corresponds to three times the mass resolution around 
its known mass. 
The $\Lc$ candidate is then combined with two (three) additional charged particles
  to form a $\Xiccp\;(\Xiccpp)$ candidate.
These additional particles must form a good-quality vertex with the $\Lc$ candidate, and this vertex is required to be upstream of the $\Lc$ decay vertex.
Each $\PXi_{cc}^{+(+)}$ candidate is required to have
a transverse momentum $\pt>3\gevc$ and to
be consistent with originating from its associated PV,
defined as 
the PV with respect to which 
it has the smallest impact parameter (IP). 
To avoid candidates including duplicated segments of tracks, 
all track pairs are required to have an opening angle larger than 0.7~mrad.

In order to improve the signal purity, multivariate classifiers are trained to separate signal from background. 
First, a classifier based on a boosted decision tree (BDT)  algorithm~\cite{Breiman,AdaBoost} implemented in the TMVA
toolkit~\cite{Hocker:2007ht,*TMVA4} is developed to select $\Lc$ candidates. To mitigate the effects of the slightly different distributions of particle-identification variables in data and simulation, the classifier is trained directly on data. The signal proxy consists of $\Lc$ candidates from a subset of data, with the background statistically subtracted using the \sPlot technique~\cite{Pivk:2004ty}, while candidates from the $\Lc$ mass sidebands are used as a background proxy. 
The $\Lc$ BDT response  
is then included in the training of a 
second BDT classifier used for the $\PXi_{cc}^{+(+)}$ decay selection. 
It uses simulated signal events as a signal proxy and wrong-sign (WS) $\Lc\Km\pim$ combinations in the mass range 3600--3640\mevcc in data as a background proxy. Both the signal and  background proxies are required to 
satisfy all previous selection requirements.
Variables 
used in the BDT training include 
the $\Xiccp$  decay-vertex fit quality,
the $\Xiccp$ IP significance,
the angle between the $\Xiccp$ momentum and displacement vector from the associated PV,
and the flight-distance \chisq between the PV and the decay vertex of the $\Xiccp$ candidate. 
The flight-distance $\chisq$ is defined as the $\chisq$ of the hypothesis that 
the decay vertex of the candidate coincides with its associated PV.
Variables associated with the decay products  of the $\Xiccp$ candidates used in the BDT training include their \pt and IP significance,
the vertex-fit quality of the \Lc candidates, and the \pt and IP significance of the \Lc decay products. 
Particle-identification information of the final-state particles 
from the $\Xiccp$ decay 
is also used.

The requirement applied to the classifier response is determined by 
maximizing the expected
value of the figure of merit 
$\varepsilon/(\frac{5}{2}+\sqrt{B})$~\cite{Punzi:2003bu},
where $\varepsilon$ is the estimated signal efficiency from simulation and
$B$ is the number of WS background candidates within $\pm 2\sigma_m$ from the expected signal mass. 
Here, $\sigma_m = 5.36\mevcc$ is the mass resolution, obtained from simulation and corrected for a 3\% data-simulation difference measured with the $\Xiccpp$ control sample.
The performance of the multivariate selection is also tested
on simulated signal candidates under other lifetime hypotheses 
and is found to have no strong dependence on this input. 

The multivariate classifier
for the \decay{\Xiccpp}{\Lc\Km\pip\pip} decay is developed following the same strategy as that for the \decay{\Xiccp}{\Lc\Km\pip} decay,
except that the threshold of the classifier response is determined by maximizing the signal significance $S/\sqrt{S+B}$, 
where $S$ is the expected signal yield,
and $B$ is the expected number of background candidates under the signal peak.

After the selection, an event may contain more than one $\PXi_{cc}^{+(+)}$ candidate.
According to studies on simulated decays and the WS control sample, multiple candidates in the same event 
have very similar $\PXi_{cc}^{+(+)}$ mass 
if they are obtained from the same final-state
tracks, but with two such tracks exchanged (\eg the \Km from the \Lc decay and 
the \Km from the $\PXi_{cc}^{+(+)}$ decay).
In this case,
one candidate is chosen arbitrarily.
Other kinds of multiple candidates, which account for 2\% ($7$\%) of the 
selected $\Xiccp\;(\Xiccpp)$ candidates in the mass fit window, 
are not removed~\cite{Koppenburg:2017zsh} since they do not form a peaking background.

To improve the mass resolution, the invariant mass 
of $\PXi_{cc}^{+(+)}$ candidates is computed as
\begin{equation}
    \label{eq:dm}
\begin{aligned}
    m_{\rm cand}(\PXi_{cc}^{+(+)}) &= m(\Lc\Km\pip(\pip))-m([p\Km\pip]_{\Lc})+M_{\rm PDG}(\Lc), 
\end{aligned}
\end{equation}
where $m(\Lc\Km\pip(\pip))$ is the reconstructed $\PXi_{cc}^{+(+)}$ mass; $m([p\Km\pip]_{\Lc})$ is the reconstructed $\Lc$ mass; and $M_{\rm PDG}(\Lc)$ is the known \Lc mass, $2286.46 \pm 0.14 \mevcc$~\cite{BaBar:2005wur}.

\begin{figure}[t]
\begin{center}
    \includegraphics*[width=.49\textwidth]{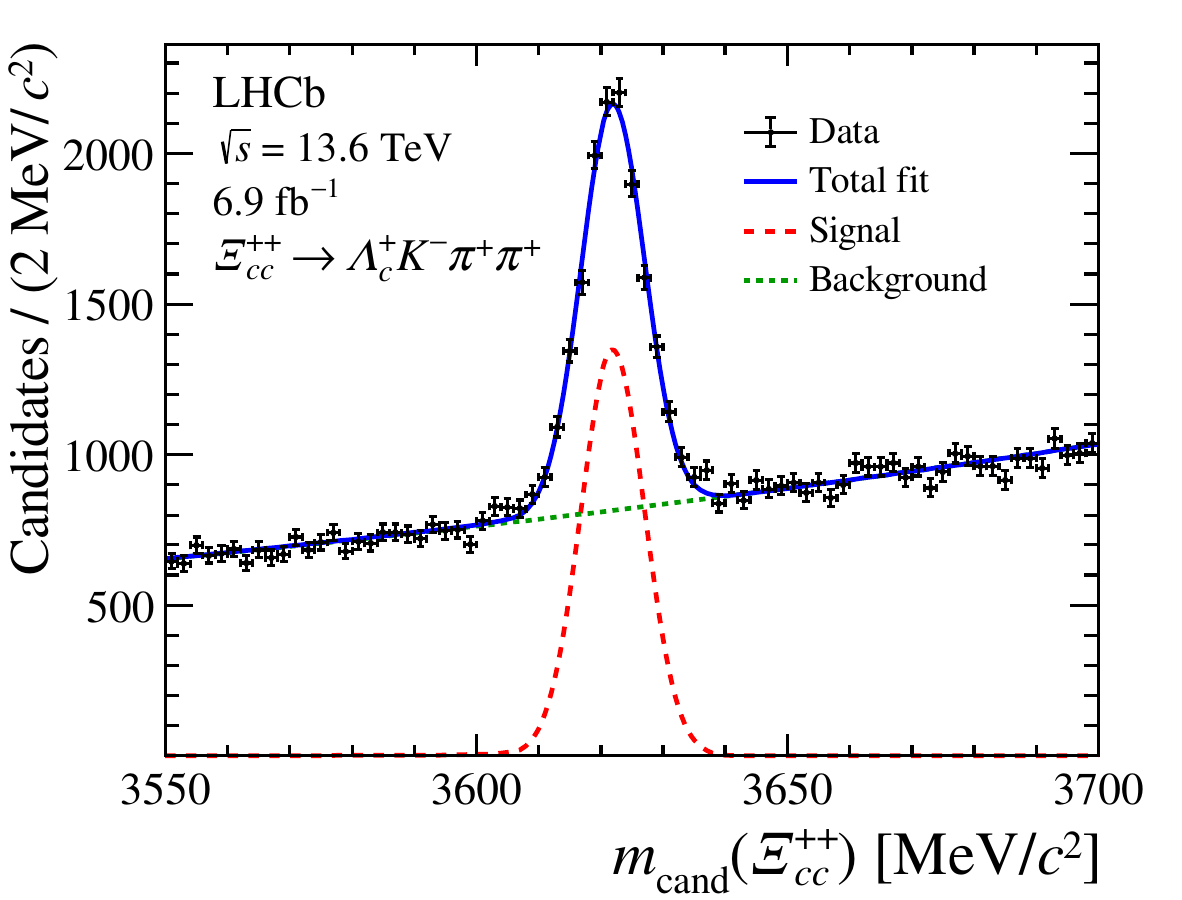}
\end{center}
\caption{\small
  Mass distribution of the selected $\Xiccpp$ 
  candidates in data, with fit results shown.}
\label{fig:XiccPP2024}
\end{figure}

The $m_{\rm cand}(\Xiccpp)$ distribution of the selected $\Xiccpp$ candidates
in data is shown in Fig.~\ref{fig:XiccPP2024}. 
An unbinned extended maximum-likelihood fit to the distribution is
performed as in Ref.~\cite{LHCb-PAPER-2019-037}, with
the signal described by a Crystal Ball
function~\cite{Skwarnicki:1986xj}, and the background parameterized by 
an exponential function.
The fitted signal yield is 
$8712 \pm 160$,
while the fitted mass 
is 
$3621.93 \pm 0.10\mevcc$,
where the uncertainties are statistical only. 
The $\Xiccpp$ signal yield per $\invfb$ of data 
is 1262 in 2024, compared with 285 in the Run~2 data~\cite{LHCb-PAPER-2019-037},
which means that the efficiency in 2024 has increased by a factor of about four compared with Run~2.

Multiple scattering can alter the values of the reconstructed mass and flight distance of the $\PXi_{cc}^{+(+)}$ candidates. 
Since the selection is more efficient for candidates with larger reconstructed flight distances, and the flight distance is correlated with the mass by the effect of multiple scattering, this can bias the $\PXi_{cc}^{+(+)}$ mass measurement. This effect is studied with charmed hadrons, 
and is found to be well reproduced by simulation~\cite{LHCb-PAPER-2017-018,LHCb-PAPER-2019-037}. 
The corresponding corrections of $-0.26\pm0.03\mevcc$ for the $\Xiccpp$ mass and $-2.0\pm 0.1\mevcc$ for the $\Xiccp$ mass (assuming a lifetime of 45\fs) are determined using simulated samples
by comparing the fitted mass of signal candidates before and after applying the selection. 
Final-state photon radiation also causes a bias in the measured mass.
It is determined with the simulation~\cite{davidson2015photos}
to be 
$-0.07 \pm  0.10\mevcc$ 
and 
$-0.14 \pm 0.10\mevcc$ 
for the $\Xiccpp$~\cite{LHCb-PAPER-2019-037} and 
$\Xiccp$ baryons, respectively. 
These corrections are applied to the fitted mass values. The uncertainties are due to the limited size of simulated samples,
and are taken as systematic uncertainties on the $\PXi_{cc}^{+(+)}$ mass.
After correcting for the bias due to 
the event selection and 
final-state radiation, 
the $\Xiccpp$ mass is determined to be 
$3621.74 \pm 0.10\mevcc$, where the uncertainty is statistical only.
This is 
consistent with the $\Xiccpp$ mass obtained from LHCb Run 1--2 data,
$3621.6 \pm 0.4\mevcc$~\cite{LHCb-PAPER-2019-037}. 

\begin{figure}
\begin{center}
    \includegraphics[width=.49\textwidth]{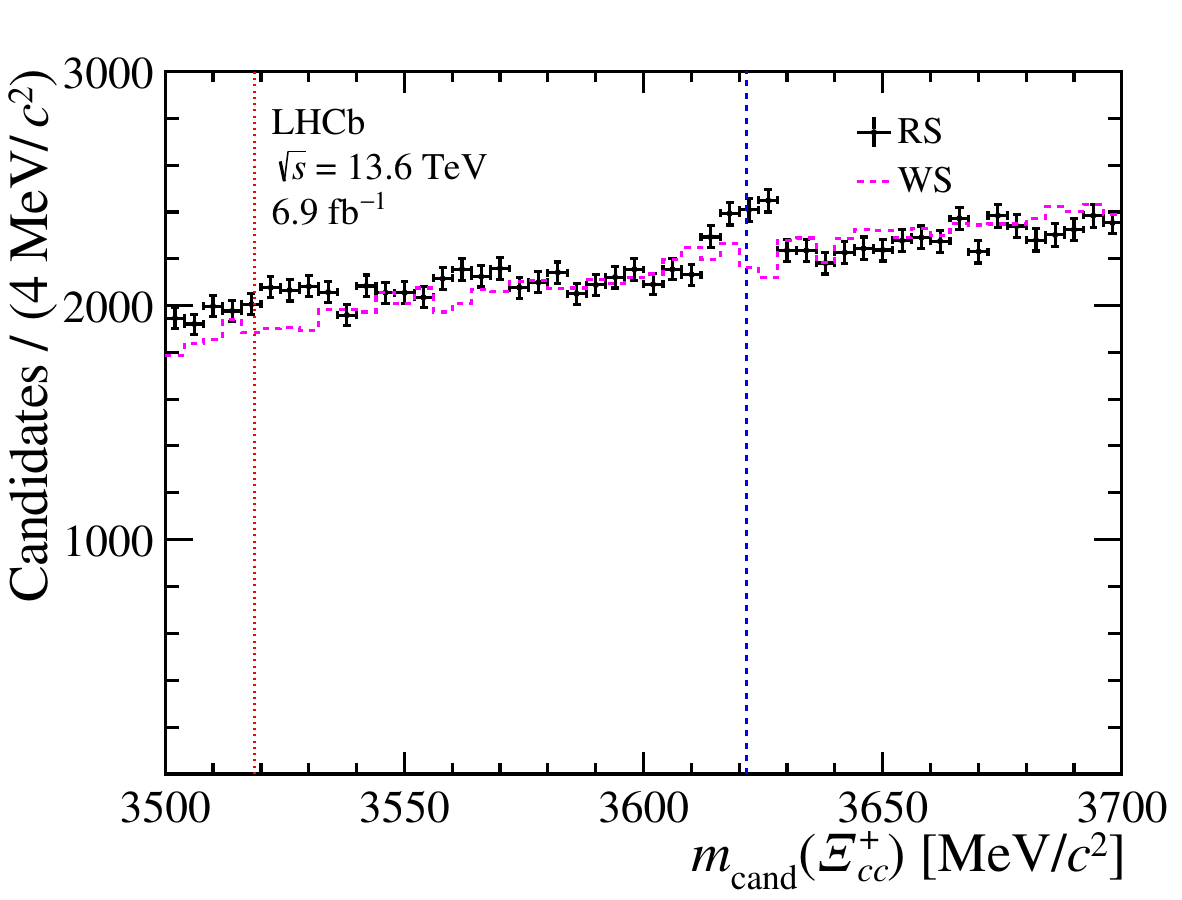}
\end{center}
\caption{\small
Mass distributions of the selected \Xiccp candidates. 
  The RS $m(\Lc\Km\pip)$ distribution is shown,
  along with the WS $m(\Lc\Km\pim)$ distribution.
  The dotted red line at $3518.7\mevcc$ indicates the mass of the \Xiccp baryon claimed by SELEX~\cite{Mattson:2002vu,Ocherashvili:2004hi} and the dashed blue line at $3621.6\mevcc$ indicates the mass of the isospin partner, the \Xiccpp baryon~\cite{LHCb-PAPER-2019-037}.
  }
\label{fig:XiccP}
\end{figure}

\begin{figure}
  \begin{center}
    \includegraphics[width=.49\textwidth]{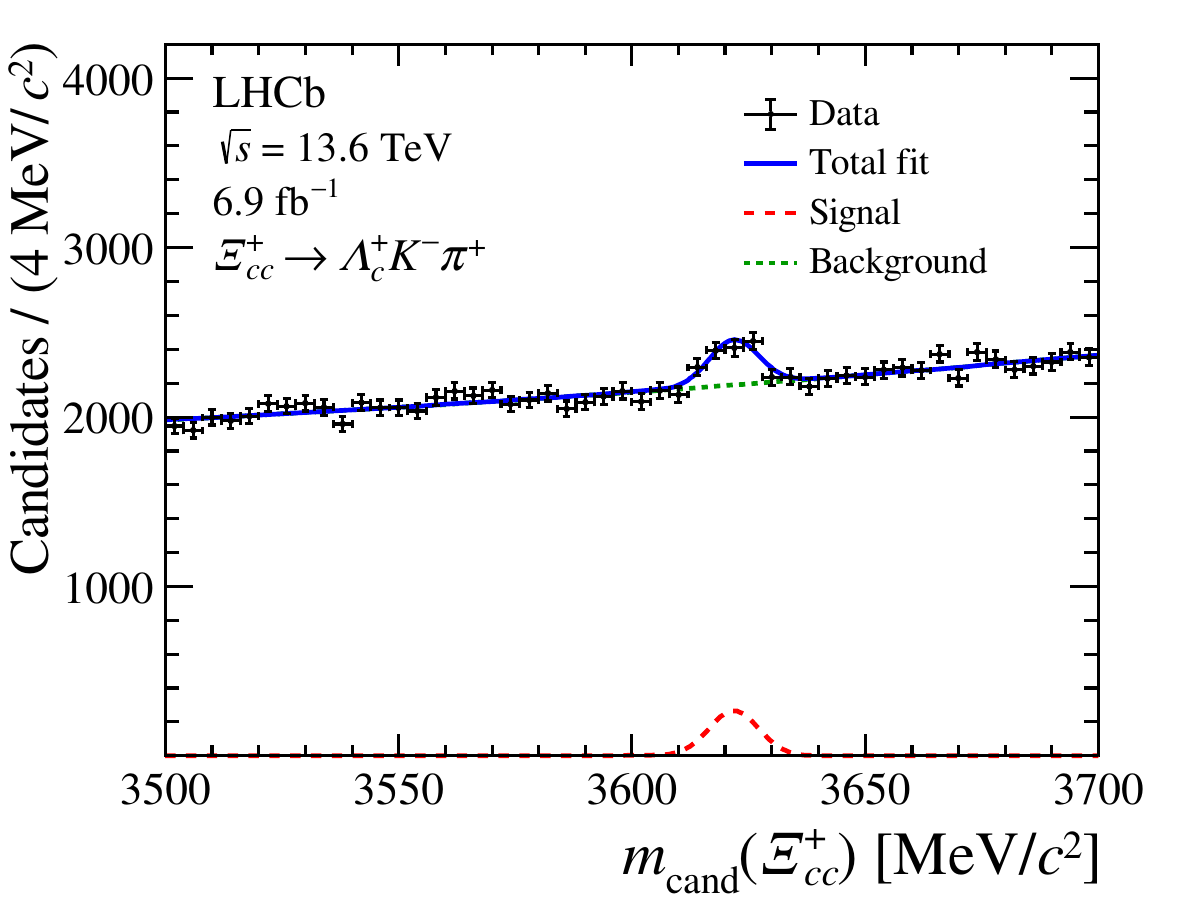}
  \end{center}
  \caption{
    Mass
    distribution of the selected $\Xiccp$ candidates in data, with fit results shown.}
  \label{fig:massFit}
\end{figure}

The $m_{\rm cand}(\Xiccp)$ distributions of the selected $\Xiccp$ candidates
for the right-sign (RS) and WS data samples
are shown in Fig.~\ref{fig:XiccP}.
A structure is visible in the signal mode at a mass of approximately 3620\mevcc. 
No significant structure is visible in the WS control sample.
To measure the properties of the structure, an unbinned extended maximum-likelihood fit is performed
to the mass distribution in the $\Lc\Km\pip$ mass window of 3500--3700\mevcc (Fig.~\ref{fig:massFit}).
This choice excludes the partially reconstructed background, \eg from \decay{\Xiccpp}{\Lc\Km\pip\pip} decays with one of the pions from the $\Xiccpp$ decay  missing. 
The peaking structure is described by a Crystal Ball function
with all parameters fixed from simulation
with the exception of the peak position and the yield.
The mass resolution is fixed to $5.36\mevcc$ as obtained from simulation and corrected for a 3\% data-simulation difference measured with the $\Xiccpp$ control sample. 
The background is described by a second-order polynomial function with parameters free to vary in the fit.
The signal yield is measured to be
$915 \pm 120$, corresponding to a local statistical significance in excess of seven standard deviations ($\sigma$) when evaluated with a likelihood-ratio test for rejection of the background-only hypothesis~\cite{Wilks:1938dza,Cowan:2010js}. 

The same structure is also seen in a $pp$ data sample collected by LHCb at $\sqrt{s}=13\tev$ in 2016--2018 data (part of Run~2), as shown in the End Matter. 
The signal yield in that sample is measured to be $304\pm 80$, 
corresponding to a statistical significance of $4\sigma$, and the mass value is consistent with that obtained in the data sample used for this analysis. 
Compared with Run~2,  
the signal yield of the \decay{\Xiccp}{\Lc\Km\pip} decay
normalized by the integrated luminosity in 2024 is only larger
by a factor of about 2.5, instead of 4 for the $\Xiccpp$ baryon. This is mainly due to the smaller lifetime of the $\Xiccp$ baryon, therefore the event-selection relies more on the momenta instead of the lifetime-related variables, and is less affected by the momentum thresholds in the hardware trigger of Run 2.

Additional cross-checks are performed to test the robustness
of the signal. These include
changing the background lineshape to an exponential function, letting the mass resolution float freely, and 
  using an alternative multivariate classifier for $\Xiccp$ without including  
  particle-identification information in the training.
In each of the tests, the significance of the structure in the
$\Lc \Km \pip$ final state remains above $6\sigma$.

The sources of systematic uncertainty affecting the measurement of the $\Xiccp$ mass, $M(\Xiccp)$, 
and the mass difference between the $\Xiccp$ and $\Xiccpp$ baryons, \mbox{$\Delta M\equiv M(\Xiccp)-M(\Xiccpp)$},
include the effects of 
the momentum-scale calibration, 
the energy loss~\cite{Chu:2025qyv}, 
the selection, 
the invariant-mass fit model,
the uncertainty on the \Lc mass,
and the unknown $\Xiccp$ lifetime.
They are summarized in Table~\ref{tab:MassMeasurementSys}.
After the momentum-scale calibration, an uncertainty 
of $\pm 0.02\%$ is assigned, which is estimated by using samples of 
$\PUpsilon$ and $\psi(2S)$ decays. 
The energy loss is evaluated following Ref.~\cite{Chu:2025qyv} 
by studying the fitted mass of \decay{\psi(2S)}{\jpsi\pip\pim} decays
as a function of the momentum of the pair of pions 
to determine the bias.
The bias on the \Xiccp mass due to the selection is estimated using simulation, and has a uncertainty due to the limited size of simulation sample. 
Such bias depends on the unknown \Xiccp lifetime,
which introduces a further source of uncertainty on the correction.
This uncertainty is estimated by repeating the analysis for other \Xiccp lifetime hypotheses
between 15 and 160$\fs$ (adding approximately 20$\fs$ on both sides of the predicted range 36--138$\fs$).
The bias is estimated to be $+3.3 \pm 0.2 \mevcc$ 
for $15\fs$
and $+0.1\pm 0.1\mevcc$ for $160\fs$, respectively.
These correspond to an additional correction of $-1.3\mevcc$ for $15\fs$
and $+1.9\mevcc$ for $160\fs$, and are listed separately. 
The dependence of the measurement on the mass fit model is estimated by
varying the shape parameters that are fixed according to simulation,
by using alternative signal and background models (a double Crystal Ball function as signal or an exponential function as background),
and by repeating the fits in different mass ranges.
The largest deviation in the results
is assigned as a systematic uncertainty.
Finally, since the \Xiccp mass is measured relative to the \Lc mass,
the uncertainty of $0.14$\mevcc on the world-average value~\cite{BaBar:2005wur} of the latter is included.
The total systematic uncertainty is obtained as the sum in quadrature of these uncertainties.

\begin{table}[tb]
    \caption{Systematic uncertainties on $M(\Xiccp)$ and the mass difference \mbox{$\Delta M\equiv M(\Xiccp)-M(\Xiccpp)$}. }
\centering
\begin{tabular}{lrr}
\hline
Source                       & $M(\Xiccp)$ [$\mevcc$] & $\Delta M$ [$\mevcc$] \\
\hline
Momentum-scale calibration   & 0.14 & 0.03 \\
Energy loss                  & 0.10 & 0.05 \\
Selection bias correction    & 0.10 & 0.10 \\
Mass fit model               & 0.10 & 0.10 \\
$\Lc$ mass uncertainty       & 0.14 & --\phantom{x}\\
\hline
Sum in quadrature            & 0.26 & 0.15 \\
\hline
\rule{0pt}{12pt}\hskip-0.5pt
$\Xiccp$ lifetime      & $_{-1.30}^{+1.90}$  & $_{-1.30}^{+1.90}$ \\
\hline
\end{tabular}\label{tab:MassMeasurementSys}
\end{table}

In summary, a significant structure 
that is consistent with the $\Xiccp$ baryon 
is observed in the final state $\Lc \Km \pip$ 
in a $pp$ data sample collected with the \lhcb Run~3 detector during 2024 at  
a center-of-mass energy of 13.6\tev,
with a signal yield of $915\pm 120$.
The mass of the structure is measured to be 
\begin{equation}
    M(\Xiccp)=\Xiccmass,\nonumber
\end{equation}
where the first uncertainty is statistical, the second is systematic, and the third 
is due to the unknown $\Xiccp$ lifetime, 
  which is assumed to lie in the range 15--160$\fs$ with a baseline value of 45\fs.
A measurement of the $\Xiccp$ lifetime would   
significantly reduce the latter uncertainty. 
The mass difference between the \Xiccp and $\Xiccpp$ states is determined to be 
\begin{equation}
    \Delta M=\DeltaMass, \nonumber
\end{equation}
which is consistent with theoretical predictions~\cite{Hwang:2008dj,
Brodsky:2011zs,
BMW:2014pzb,
Wei:2015gsa,
Karliner:2017gml} that include partial cancellation of QCD and QED effects.
The mass of the observed \Xiccp state is greater
than that of the \Xiccp state claimed by the SELEX collaboration~\cite{Mattson:2002vu,Ocherashvili:2004hi}
by $101 \pm 2$\mevcc, and consistent with the value corresponding to the local excess found in previous LHCb searches based on the Run~1 and 2 samples~\cite{LHCb-PAPER-2019-029,LHCb-PAPER-2021-019}.
Consequently, 
the interpretation of the SELEX state as the \Xiccp baryon is not supported.

This observation was made possible thanks to a significant increase in the selection efficiency for this channel in the 2024 data sample compared with Run 2, stemming from the enhanced capabilities of the upgraded LHCb detector and its trigger system.

\section*{Acknowledgements}
%
%
\noindent We express our gratitude to our colleagues in the CERN
accelerator departments for the excellent performance of the LHC. We
thank the technical and administrative staff at the LHCb
institutes.
We acknowledge support from CERN and from the national agencies:
ARC (Australia);
CAPES, CNPq, FAPERJ and FINEP (Brazil); 
MOST and NSFC (China); 
CNRS/IN2P3 and CEA (France);  
BMFTR, DFG and MPG (Germany);
INFN (Italy); 
NWO (Netherlands); 
MNiSW and NCN (Poland); 
MEC/IFA (Romania); 
MICIU and AEI (Spain);
SNSF and SER (Switzerland); 
NASU (Ukraine); 
STFC (United Kingdom); 
DOE NP and NSF (USA).
We acknowledge the computing resources that are provided by ARDC (Australia), 
CBPF (Brazil),
CERN, 
IHEP and LZU (China),
IN2P3 (France), 
KIT and DESY (Germany), 
INFN (Italy), 
SURF (Netherlands),
Polish WLCG (Poland),
IFIN-HH (Romania), 
PIC (Spain), CSCS (Switzerland), 
GridPP (United Kingdom),
and NSF (USA).  
We are indebted to the communities behind the multiple open-source
software packages on which we depend.
Individual groups or members have received support from
RTP (Australia), 
Key Research Program of Frontier Sciences of CAS, CAS PIFI, CAS CCEPP (China); 
Minciencias (Colombia);
EPLANET, Marie Sk\l{}odowska-Curie Actions, ERC and NextGenerationEU (European Union);
A*MIDEX, ANR, IPhU and Labex P2IO, and R\'{e}gion Auvergne-Rh\^{o}ne-Alpes (France);
Alexander-von-Humboldt Foundation (Germany);
ICSC (Italy); 
Severo Ochoa and Mar\'ia de Maeztu Units of Excellence, GVA, XuntaGal, GENCAT, InTalent-Inditex and Prog.~Atracci\'on Talento CM (Spain);
the Leverhulme Trust, the Royal Society and UKRI (United Kingdom).

\addcontentsline{toc}{section}{References}
\bibliographystyle{LHCb/LHCb}
\bibliography{main,xicc-refs,LHCb/standard,LHCb/LHCb-PAPER,LHCb/LHCb-DP,LHCb/LHCb-TDR}

\clearpage
\clearpage

\section*{End matter}
\label{sec:EndMatter}

The Letter describes the observation of a narrow structure in the $\Lc \Km \pip$ mass
spectrum in a sample of proton-proton data collected by the LHCb Run~3 detector in 2024 at a center-of-mass energy of 13.6\tev, corresponding to an integrated luminosity of $\totLumi$. 
As an additional cross-check, 
the Run~2 (2016--2018) data sample recorded by the LHCb experiment
at a center-of-mass energy of 13\tev 
as used in the previous search~\cite{LHCb-PAPER-2019-029}
has been reanalyzed using a previously unexplored data stream
that selects \decay{\Lc}{\proton\Km\pip} inclusively.
With this data stream the transverse momentum requirements on the kaon and pion from the $\Xiccp$ decay 
can be relaxed from 500\mevc to 200\mevc. 
An improved selection has been developed 
by including the particle-identification information in the multivariate analysis
and taking into account that the $\Xiccp$ lifetime was recently estimated to be about $45\fs$~\cite{Cheng:2018mwu,Dulibic:2023jeu}, which is shorter than the assumed $80\fs$ in Ref.~\cite{LHCb-PAPER-2019-029}. 

\begin{figure}[!b]
  \begin{center}
    \includegraphics[width=.49\textwidth]{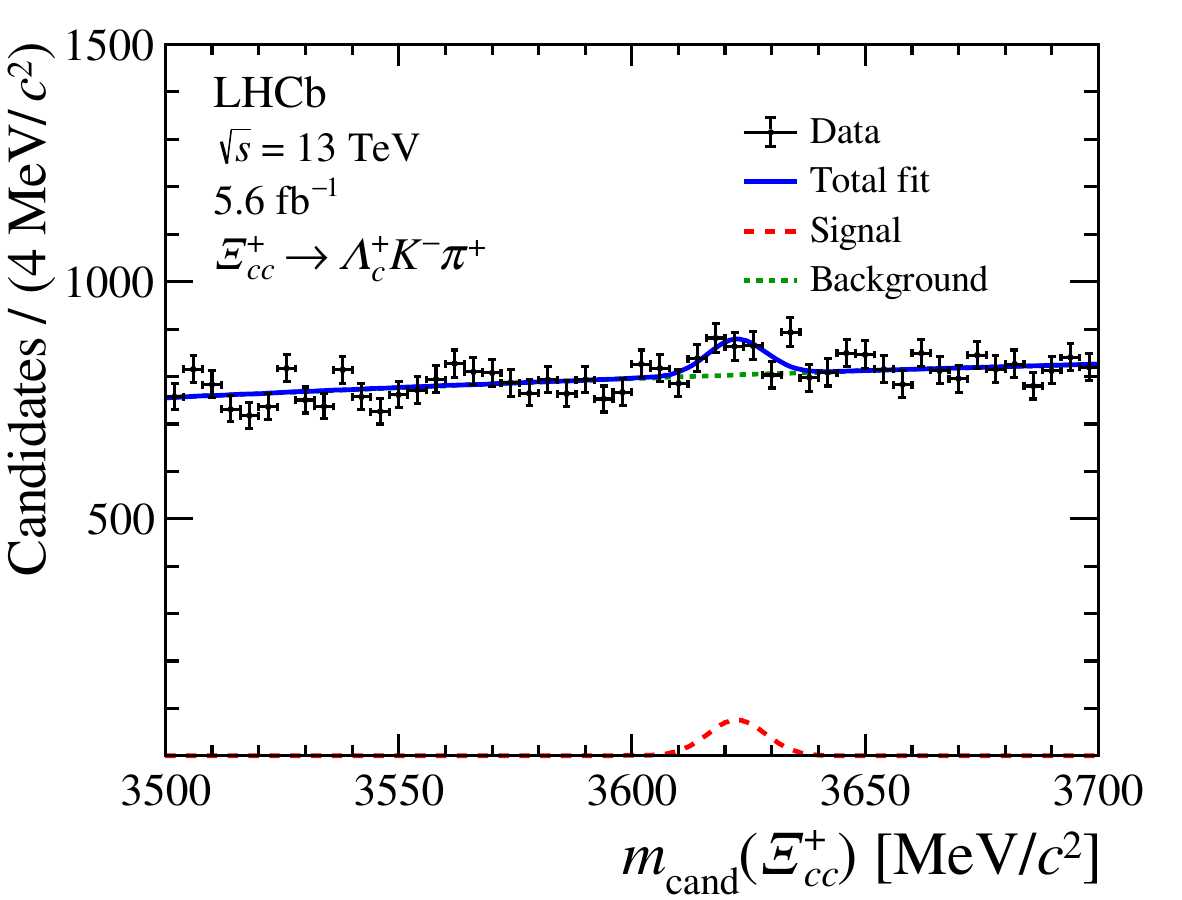}
  \end{center}
  \vspace*{-.7cm}
  \caption{
    Mass
    distribution of the selected $\Xiccp$ candidates in Run~2 data, with the result of the fit also shown.}
  \label{fig:massFitRun2}
\end{figure}

Figure~\ref{fig:massFitRun2} shows the $\Xiccp$ mass spectrum 
in the Run~2 sample after the final selection. 
The fit procedure described in the Letter is applied to 
the selected candidates, with the mass resolution fixed to 6.28\mevcc as determined from simulation after correcting a difference of 10\% between simulation and data found for the $\Xiccpp$ baryon. 
The signal yield is measured to be $304\pm 80$, and 
corresponds to a statistical significance of $4\sigma$. The $\Xiccp$ mass is determined to be $3620.8 \pm 2.2\mevcc$, where the uncertainty is statistical only. This is consistent with the $\Xiccp$ mass determined using 2024 data, 
$\Xiccmass$.

\clearpage

\centerline
{\large\bf LHCb collaboration}
\begin
{flushleft}
\small
R.~Aaij$^{38}$\lhcborcid{0000-0003-0533-1952},
M. ~Abdelfatah$^{69}$,
A.S.W.~Abdelmotteleb$^{57}$\lhcborcid{0000-0001-7905-0542},
C.~Abellan~Beteta$^{51}$\lhcborcid{0009-0009-0869-6798},
F.~Abudin\'en$^{59}$\lhcborcid{0000-0002-6737-3528},
T.~Ackernley$^{61}$\lhcborcid{0000-0002-5951-3498},
A. A. ~Adefisoye$^{69}$\lhcborcid{0000-0003-2448-1550},
B.~Adeva$^{47}$\lhcborcid{0000-0001-9756-3712},
M.~Adinolfi$^{55}$\lhcborcid{0000-0002-1326-1264},
P.~Adlarson$^{87,42}$\lhcborcid{0000-0001-6280-3851}
% ...

{\footnotesize \it

% ...
$^{38}$Nikhef National Institute for Subatomic Physics, Amsterdam, Netherlands\\
% ...
$^{42}$National Center for Nuclear Research (NCBJ), Warsaw, Poland\\
% ...
$^{47}$Instituto Galego de F{\'\i}sica de Altas Enerx{\'\i}as (IGFAE), Universidade de Santiago de Compostela, Santiago de Compostela, Spain\\
% ...
$^{51}$Physik-Institut, Universit{\"a}t Z{\"u}rich, Z{\"u}rich, Switzerland\\
% ...
$^{55}$H.H. Wills Physics Laboratory, University of Bristol, Bristol, United Kingdom\\
% ...
$^{57}$Department of Physics, University of Warwick, Coventry, United Kingdom\\
% ...
$^{59}$School of Physics and Astronomy, University of Edinburgh, Edinburgh, United Kingdom\\
$^{60}$School of Physics and Astronomy, University of Glasgow, Glasgow, United Kingdom\\
$^{61}$Oliver Lodge Laboratory, University of Liverpool, Liverpool, United Kingdom\\
% ...
$^{69}$Syracuse University, Syracuse, NY, United States\\
% ...
$^{87}$Department of Physics and Astronomy, Uppsala University, Uppsala, Sweden, associated to $^{60}$\\
% ...
}
\end{flushleft}

\end{document}